\documentclass[superscriptaddress,twocolumn,10pt,prx,floatfix]{revtex4-2}
\usepackage[sort&compress]{natbib}

\input{glyphtounicode}
\usepackage[T1]{fontenc}
\usepackage[utf8]{inputenc}
\usepackage[english]{babel}

\usepackage{amsmath}
\usepackage{amsfonts}
\usepackage{amssymb}
\usepackage{braket}
\usepackage{dsfont}
\usepackage{xfrac}

\usepackage{graphicx}
\graphicspath{{./}}
\usepackage{float}

\usepackage[pdftex,
    bookmarks=true,
    bookmarksopen=true,
    bookmarksnumbered=true,
    colorlinks=true,
    linkcolor=black,
    citecolor=black,
    urlcolor=black
]{hyperref}

\usepackage{tabularx}
\usepackage{array}
\usepackage[table]{xcolor}
\usepackage{booktabs}
\usepackage{multirow}
\usepackage{caption}
\usepackage[shortlabels]{enumitem}
\usepackage{hhline}
\usepackage{makecell}
\usepackage{placeins}

\usepackage{siunitx}
\emergencystretch=\maxdimen
\allowdisplaybreaks

\usepackage{natbib}

\usepackage{comment}
\DeclareCaptionLabelSeparator{line}{\hspace{2pt}\bf{|}\hspace{2pt}}
\begin{document}

\title{Multi-Qubit Entanglement of Unit Cell Pairs in SiMOS}
\author{Cameron Jones}
\email{cameron.jones@unsw.edu.au}
\affiliation{School of Electrical Engineering and Telecommunications, University of New South Wales, Sydney, NSW, Australia}

\author{Jonathan Y. Huang}
\affiliation{School of Electrical Engineering and Telecommunications, University of New South Wales, Sydney, NSW, Australia}

\author{Santiago Serrano}
\affiliation{School of Electrical Engineering and Telecommunications, University of New South Wales, Sydney, NSW, Australia}
\affiliation{Diraq Pty Ltd, Sydney, NSW, Australia}

\author{MengKe Feng}
\affiliation{School of Electrical Engineering and Telecommunications, University of New South Wales, Sydney, NSW, Australia}
\affiliation{Diraq Pty Ltd, Sydney, NSW, Australia}

\author{Gerardo A. Paz-Silva}
\affiliation{School of Electrical Engineering and Telecommunications, University of New South Wales, Sydney, NSW, Australia}
\affiliation{Diraq Pty Ltd, Sydney, NSW, Australia}

\author{Tuomo Tanttu}
\affiliation{School of Electrical Engineering and Telecommunications, University of New South Wales, Sydney, NSW, Australia}
\affiliation{Diraq Pty Ltd, Sydney, NSW, Australia}

\author{Paul Steinacker}
\affiliation{School of Electrical Engineering and Telecommunications, University of New South Wales, Sydney, NSW, Australia}
\affiliation{Diraq Pty Ltd, Sydney, NSW, Australia}

\author{Fay E. Hudson}
\affiliation{School of Electrical Engineering and Telecommunications, University of New South Wales, Sydney, NSW, Australia}
\affiliation{Diraq Pty Ltd, Sydney, NSW, Australia}

\author{Wee Han Lim}
\affiliation{School of Electrical Engineering and Telecommunications, University of New South Wales, Sydney, NSW, Australia}
\affiliation{Diraq Pty Ltd, Sydney, NSW, Australia}

\author{Nikolay V. Abrosimov}
\affiliation{Leibniz-Institut für Kristallzüchtung, Berlin, Germany}

\author{Hans-Joachim Pohl}
\affiliation{VITCON Projectconsult GmbH, Jena, Germany}

\author{Michael L. W. Thewalt}
\affiliation{Department of Physics, Simon Fraser University, Vancouver, British Columbia, Canada}

\author{Andrew S. Dzurak}
\affiliation{School of Electrical Engineering and Telecommunications, University of New South Wales, Sydney, NSW, Australia}
\affiliation{Diraq Pty Ltd, Sydney, NSW, Australia}

\author{Andre Saraiva}
\affiliation{School of Electrical Engineering and Telecommunications, University of New South Wales, Sydney, NSW, Australia}
\affiliation{Diraq Pty Ltd, Sydney, NSW, Australia}

\author{Arne Laucht}
\affiliation{School of Electrical Engineering and Telecommunications, University of New South Wales, Sydney, NSW, Australia}
\affiliation{Diraq Pty Ltd, Sydney, NSW, Australia}

\author{Chih Hwan Yang}
\email{henry.yang@unsw.edu.au}
\affiliation{School of Electrical Engineering and Telecommunications, University of New South Wales, Sydney, NSW, Australia}
\affiliation{Diraq Pty Ltd, Sydney, NSW, Australia}

\date{\today}

\begin{abstract}
Spin qubits in silicon-MOS (SiMOS) quantum dots have recently demonstrated compatibility with existing industry standard CMOS fabrication techniques. These devices have routinely achieved single- and two-qubit gate fidelities above \SI{99}{\percent} and demonstrated highly entangled two-qubit Bell states in isolated double quantum dot (DQD) unit cells, however coupling between unit cells has remained challenging. In this work, we present a two unit cell, four-qubit SiMOS processor with universal controllability and fully parallelised state initialisation and readout. We use this processor to generate maximally entangled three-qubit states, including the Greenberger–Horne–Zeilinger (GHZ) state, and certify multipartite entanglement through violation of the classical Mermin‑witness bound. By using a fully symmetric dynamically decoupled gate sequence to create our entangled states, we are able to preserve the lifetime of the entanglement beyond $T_2^*$, to a time limited instead by $T_2^\textrm{Hahn}$. These demonstrations pave a road to the scalable operation of larger SiMOS processors, and achieving high purity, long-lived multi-qubit entangled states in them. 
\end{abstract}

\maketitle

\begin{figure*}[ht!]
    \includegraphics[width=\textwidth]{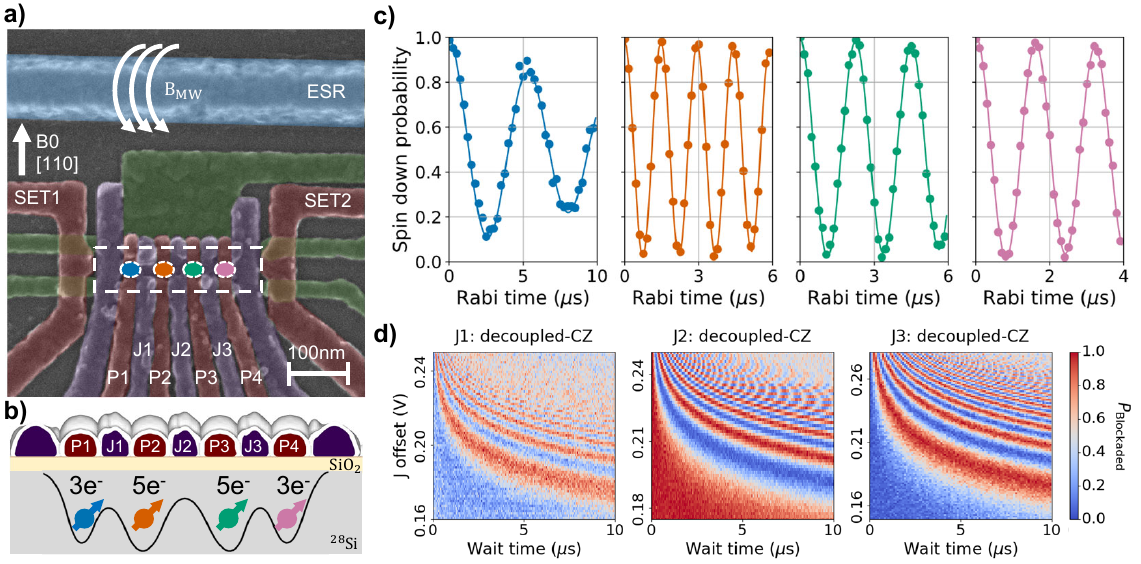}
    \caption{\textbf{Device overview and controllability.}
    \textbf{a,} A false coloured scanning electron microscope image of a device similar to that used in this work. Green (barrier gates) are metallization layer 1, red (plunger gates) are layer 2, purple (J gates) are layer 3, and blue (ESR antenna) are layer 4. 
    \textbf{b,} Schematic cross-section of the device in the dotted region of (a). Electrons are accumulated at the Si/SiOx interface under the plunger (P) gates. The device is operated in the (3,5,5,3) charge configuration.  
    \textbf{c,} Single-qubit Rabi oscillations of the qubits $\textrm{Q}_1$-$\textrm{Q}_4$ corresponding to qubits under gates P1-P4 as a function of applied microwave pulse duration.
    \textbf{d,} Decoupled controlled phase oscillations between $\textrm{Q}_1$-$\textrm{Q}_2$ (J1), $\textrm{Q}_2$-$\textrm{Q}_3$ (J2), and $\textrm{Q}_3$-$\textrm{Q}_4$ (J3). J-offset corresponds to the dynamic voltage applied with reference to the operating voltage used during single-qubit gates. For fits of the exchange rate controllability see Extended Data Fig.~\ref{fig:extended_fig_3}. 
     }
    \label{fig:main_fig_1}
\end{figure*}

To realise a utility‑scale quantum computer, one must be able to fabricate quantum processors containing millions of qubits while maintaining high yield and qubit quality~\cite{gidney_how_2021, scholten_assessing_2024}. Concurrent operation of these qubits is essential, requiring that processors be tuned to regimes where qubit initialisation, control, and readout can be performed reliably across the device~\cite{divincenzo_physical_2000}, and with fidelities compatible with quantum error‑correction (QEC) codes~\cite{taylor_fault-tolerant_2005, fowler_surface_2012, google_quantum_ai_quantum_2025}. The generation of high fidelity multi‑qubit entangled states is another critical capability, as these states underpin the operation of QEC codes and enable nontrivial quantum algorithms. Demonstrating such states has therefore become a standard benchmarking tool to validate the performance of quantum processors.

Semiconductor-based spin qubits have emerged in recent decades as a promising architecture for utility‑scale quantum computation, offering long coherence times and fast, high fidelity gate operations~\cite{tyryshkin_electron_2012, veldhorst_addressable_2014, kawakami_electrical_2014, yoneda_quantum-dot_2018, noiri_fast_2022, xue_quantum_2022, tanttu_assessment_2024, madzik_operating_2025}. They have also shown the ability to operate at elevated temperatures~\cite{petit_universal_2020, yang_operation_2020, huang_high-fidelity_2024}, a major scalability advantage for the platform. Leveraging these qualities, a number of multi-qubit processors have been fabricated in silicon germanium (SiGe) heterostructures and donor based architectures, and have validated their operation by certifying entanglement between three of more qubits~\cite{hendrickx_four-qubit_2021, takeda_quantum_2021, takeda_quantum_2022, philips_universal_2022, thorvaldson_grovers_2025, edlbauer_11-qubit_2025, fernandez_de_fuentes_running_2026, undseth_weight-four_2026, dijkema_simultaneous_2026}.

Silicon metal-oxide semiconductor (SiMOS) qubits have recently demonstrated fabrication compatibility with industry‑standard CMOS manufacturing processes~\cite{zwerver_qubits_2022, elsayed_low_2024, steinacker_industry-compatible_2025, hamonic_foundry-fabricated_2025, nickl_eight-qubit_2025, chittock-wood_radio-frequency_2025}, albeit in systems of isolated double-quantum dots (DQD). The resulting improvements in material quality and processing tolerances in these devices have resulted in routine single‑ and two‑qubit gate fidelities and SPAM levels exceeding \SI{99}{\percent}~\cite{stuyck_demonstration_2024, steinacker_industry-compatible_2025}. Moreover, these devices have demonstrated integration with classical CMOS circuitry~\cite{bartee_spin-qubit_2025, thomas_rapid_2025, clarke_spin_2025}, enabling prospects for co‑integrated quantum–classical control on chip.

In this work, we build on these advances and extend the SiMOS platform to a four‑qubit processor. The device, which was fabricated in an academic cleanroom, and is operated as DQD unit cells. It is tuned such that exchange coupling can be activated between each neighbouring qubit, allowing for universal controllability. 

Using this device, we investigate several important metrics for scalability. Single‑ and two‑qubit gates are benchmarked using gate set tomography (GST)~\cite{nielsen_gate_2021}, and we measure two maximally entangled three‑qubit states, including the Greenberger–Horne–Zeilinger (GHZ) state. By designing entangling circuits in a laddered structure, we can apply optimally timed refocusing pulses to extend the lifetime of multipartite entanglement beyond $T_2^*$. The circuit design generalises to arrays with an arbitrary number of qubits, and is not impacted by varying gate times across qubits. Recognising that overall operating speed (which is dominated by state preparation and measurement) is an important metric for processor efficiency, we perform simultaneous initialisation and readout of all qubits by parallelising the control of each DQD unit cell.

\section*{Device characteristics}

Our four-qubit processor used in this work is a linear array of quantum dots in a SiMOS device, which is fabricated using isotopically enriched silicon stock with \SI{50}{ppm} residual $^{29}\textrm{Si}$. Electrons are electrostatically confined within the quantum dots under each plunger gate by applying voltages to the aluminium (Al) gates. Fig.~\ref{fig:main_fig_1}a shows a scanning electron microscope (SEM) image of a device nominally identical to the one used in this work, which has also been used in \cite{serrano_improved_2024, jones_mid-circuit_2026}.

The processor is operated in the charge configuration $(N_\textrm{P1}, N_\textrm{P2}, N_\textrm{P3}, N_\textrm{P4}) = (3,5,5,3)$ where $N_\textrm{P}$ represents the electron number in each dot (see Fig.~\ref{fig:main_fig_1}b). Qubits are formed by the unpaired electron spin in each dot, and a static magnetic field of \SI{410}{\milli\tesla} is applied in-plane to the device. This provides qubit Larmor frequencies of \SI{11.445}{\giga\hertz}, \SI{11.427}{\giga\hertz}, \SI{11.407}{\giga\hertz}, and \SI{11.422}{\giga\hertz} for qubits $\textrm{Q}_1$ to $\textrm{Q}_4$ respectively. We perform single-qubit gates by applying an RF microwave field via an on-chip Balun antenna \cite{dehollain_nanoscale_2013} on resonance with the Larmor frequency of the respective qubit. We measure Rabi frequencies for qubits $\textrm{Q}_2$--$\textrm{Q}_4$ between \SI{443}{\kilo\hertz} -- \SI{681}{\kilo\hertz},  $T_{2}^{*}$ times between \SI{4.8}{\micro\second} -- \SI{6.2}{\micro\second}, and $T_{2}^\textrm{Hahn}$ times between \SI{76.3}{\micro\second} -- \SI{87.2}{\micro\second} (see Extended Data Fig.~\ref{fig:extended_fig_2}, Table \ref{tab:GST_1Q_fid}). 

We observe a comparably slow Rabi drive on qubit $\textrm{Q}_1$, which has an overall Rabi Q-factor ($\frac{T_{\textrm{Rabi}}}{T_2^{\textrm{Rabi}}}$) of approximately 2.5, an order of magnitude lower than the other three qubits. This means that high fidelity control of this qubit cannot be realistically achieved in the current regime, despite having similar $T_{2}^\textrm{*}$ and $T_{2}^\textrm{Hahn}$ to the other qubits. We do however find that the initialisation and readout of qubit $\textrm{Q}_1$ can still be performed to a high fidelity, hence it is used primarily as an ancilla qubit.  

The inter-dot tunnel coupling between each pair of qubits is controlled via exchange (J) gates between each dot. We measure exchange coupling between each neighbouring pair as a function of J-gate voltage offset from the biasing used for single-qubit gates (Fig.~\ref{fig:extended_fig_1}d). From these measurements exchange rate controllability is calculated as 24.3(12) dec $V_\textrm{J1}$\textsuperscript{-1}, 27.9(6) dec $V_\textrm{J2}$\textsuperscript{-1} and 15.6(1) dec $V_\textrm{J3}$\textsuperscript{-1} for pairs $\textrm{Q}_1$--$\textrm{Q}_2$, $\textrm{Q}_2$--$\textrm{Q}_3$ and $\textrm{Q}_3$--$\textrm{Q}_4$ respectively (see Extended Data Fig.~\ref{fig:extended_fig_3}). This exchange controllability allows us to implement a controlled-phase operation as our native two-qubit gate. To reduce the impact of charge noise on the qubit during the exchange pulse, we use a decoupled-CZ (dCZ) gate, which has the added advantage of cancelling any phase error from J-gate-induced Stark shift \cite{tanttu_assessment_2024}. At the single-qubit operating voltage, the residual exchange rate is extrapolated to be less than a kilohertz, which is much slower than our circuit execution times. 

To complete the native gate set, we implement a virtual $\sqrt{\textrm{Z}}$ by updating the phase of the oscillator used to track the qubit Larmor frequency. This operation takes a single clock cycle of our FPGA (\SI{4}{\nano\second}), and hence is not considered to be a significant source of error, as the gate time is at least 3 orders of magnitude quicker than the relevant $T_2$ times.

Having demonstrated access to a universal qubit gate set $\{\sqrt{X}, \sqrt{Z}, \textrm{dCZ}\}$, we benchmark the performance of these gates on qubits $\textrm{Q}_2$--$\textrm{Q}_4$ using GST~\cite{nielsen_gate_2021}. The GST estimates indicate that single-qubit on-target fidelities are above \SI{99}{\percent} on all measured qubits. When these same single‑qubit gates are characterised within a two‑qubit subspace such that dephasing of an idle qubit is included, the extracted fidelities decrease to between  \SI{97.35}{\percent} and \SI{98.13}{\percent}, depending on the pair measured (see Extended Data Table \ref{tab:GST_1Q_fid}). Similarly, the extracted dCZ gate fidelities are \SI{95.82}{\percent} and \SI{95.53}{\percent} (see Extended Data Table \ref{tab:GST_2Q_fid}). 

The increased infidelity of the $\sqrt{X}$ gate when analysed in the two qubit context indicates that the spectator qubit dephasing is a dominant error source. In the GST framework, average gate infidelity is linearly sensitive to stochastic (incoherent) noise but only quadratically sensitive to coherent unitary errors \cite{blume-kohout_taxonomy_2022}. This explains why decoherence in the expanded Hilbert space dominates over any coherent miscalibration or crosstalk error. Given that the sample is isotopically purified to \SI{50}{ppm} $^{29}\textrm{Si}$, the dephasing errors are likely driven by charge noise, rather than the result of hyperfine coupling to nuclear spins in the environment. This is an encouraging conclusion, as foundry fabricated devices have been shown to achieve cleaner charge noise environments compared to academic fabrication facilities in which this device was made~\cite{elsayed_low_2024, steinacker_industry-compatible_2025}. This suggests that better gate performance can be achieved in larger SiMOS arrays by utilising foundry manufacturing.

\begin{figure*}[ht!]
    \includegraphics[width=\textwidth]{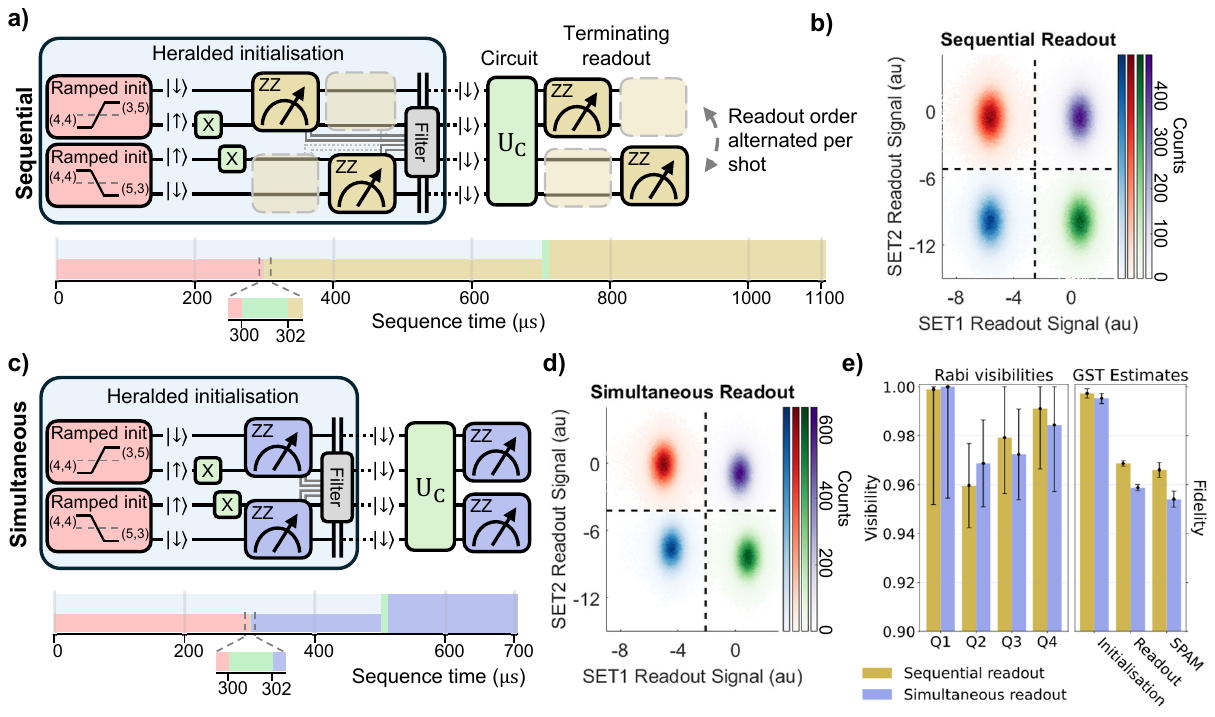}
    \caption{\textbf{Device operating sequence and readout techniques.}
    \textbf{a,} Schematic diagram showing the sequence structure for a single-shot experiment using the sequential readout technique. The four-qubit heralded initialisation uses a real-time logic state filter that re-initialises the processor unless blockaded parity is measured in both DQDs. The order of both the initialisation and terminating readouts are interchanged each measurement shot. The total measurement time is displayed below, assuming a \SI{10}{\micro\second} control operation $\textrm{U}_\textrm{c}$. 
    \textbf{b,} Two-dimensional readout histogram from GST experiments using sequential readout method. State classification thresholds are displayed as dotted black lines. See Extended Data Fig.~\ref{fig:extended_fig_5} for 1D histograms with SNR and charge fidelity fitting. 
    \textbf{c,} Sequence structure using simultaneous readout for both initialisation and terminating measurement. 
    \textbf{d,} Two-dimensional readout histogram for simultaneous readout.
    \textbf{e,} Fitted Rabi visibilities and GST estimates for initialisation, readout and combined SPAM fidelities. GST estimates are from $\textrm{Q}_2$-$\textrm{Q}_3$ GST experiment, which incorporates the initialisation and readout fidelities of both DQDs. 
     }
    \label{fig:main_fig_2}
\end{figure*}

\section*{Parallel unit cell operation}

To operate this device as a four-qubit processor, we need to reliably initialise and readout the qubit array. We treat the linear array as two DQD unit cells, P1--P2 and P3--P4, and tune the initialisation and readout protocols independently. Parity measurements via Pauli spin blockade (PSB)~\cite{seedhouse_pauli_2021} are performed in both DQDs, which are each charge coupled to a nearby radio-frequency single electron transistors (SETs)~\cite{schoelkopf_radio-frequency_1998}. A heralded initialisation via readout protocol adapted from~\cite{huang_high-fidelity_2024} and described in Methods is employed to provide robust initialisation into the spin $\ket{\downarrow\downarrow\downarrow\downarrow}$ state (computational $\ket{1111}$). This initialisation technique uses FPGA-based real-time logic to implement a state filter, which reinitialises the array of qubits if the target initial state is not measured.

In order to maintain charge stability in the DQDs, they are operated symmetrically about J2. This means during initialisation and readout, the unit cells are detuned in opposite directions, with initialisation being performed towards the (4,4,4,4)$\rightarrow$(3,5,5,3) charge transitions, and readout occurring towards (3,5,5,3)$\rightarrow$(4,4,4,4). This maintains the electrochemical potential between the P2 and P3 dots to prevent erroneous charge transitions during operation. 

To minimise circuit execution time, we implement a parallelised PSB readout protocol that enables simultaneous parity measurement of both DQDs. The tuning procedure outlined in Extended Data Fig.~\ref{fig:extended_fig_4} and described in Methods is used to identify a detuning region that supports independent parity readout of both qubit pairs while avoiding regions of cascaded readout \cite{van_diepen_electron_2021, chittock-wood_radio-frequency_2025}. Performing readout in this mode allows for the simultaneous execution of both initialisation via readout, and terminating measurements across unit cells.

For comparison, when the qubit pairs are read out sequentially (which involves measuring one DQD at a time while holding the other at the single‑qubit control point), we obtain charge‑readout fidelities at or above \SI{99.9}{\percent} in both SETs (Extended Data Fig.~\ref{fig:extended_fig_5}a) using a readout time of \SI{200}{\micro\second}. This approach doubles the time required to readout the whole qubit array, resulting in a total sequence time of roughly \SI{1.1}{\milli\second} as shown in Fig.~\ref{fig:main_fig_2}a. By contrast, the parallelised readout sequence reduces the overall time to \SI{700}{\micro\second} (Fig.~\ref{fig:main_fig_2}a,c) while maintaining readout fidelities above \SI{99.8}{\percent}. Critically, the duration of the parallel technique does not scale with the number of unit cells operated concurrently, unlike the linear scaling inherent to sequential readout.  

We further benchmark the state preparation and measurement (SPAM) error rates for both methods using GST, and by assessing correlations in the SET readouts. The 2D readout histograms from the GST experiments are plotted in Fig.~\ref{fig:main_fig_2}b,d. For the sequential case, we see no correlation between the SET1 and SET2 measurement outcomes, indicated by the vertical and horizontal alignment between modes in the 2D histogram. From this histogram we calculate an SNR of 9.4 and 6.2 for SET1 and SET2 respectively. For the simultaneous case, a slight correlation between readouts is observed. This is due to the measurement in each DQD coupling in different strengths to both SETs. We find that the degree of cross-capacitance to the furthest SET is minimal enough that the SNRs only decrease to 8.2 and 5.7 respectively, resulting in the minor reduction in charge readout fidelity mentioned previously.

The GST results estimate total SPAM to be \SI{96.6(3)}{\percent} for the sequential case, and \SI{95.4(3)}{\percent} for the simultaneous. These estimates come from a GST experiment of $\textrm{Q}_2$--$\textrm{Q}_3$, which involves the initialisation and readout of both DQDs, and are hence indicative of the SPAM rate for the full array. As verification, we measure the Rabi visibility using both readout methods, and find that the GST estimated SPAM fidelities sit within the $2\sigma$ uncertainty regions for each of the Rabi visibility fits. Furthermore, the average number of initialisation attempts to obtain the target $\ket{\downarrow\downarrow\downarrow\downarrow}$ only increases from 1.63 to 1.66 when changing from sequential to simultaneous readout.

\begin{figure*}[ht!]
    \includegraphics[width=\textwidth]{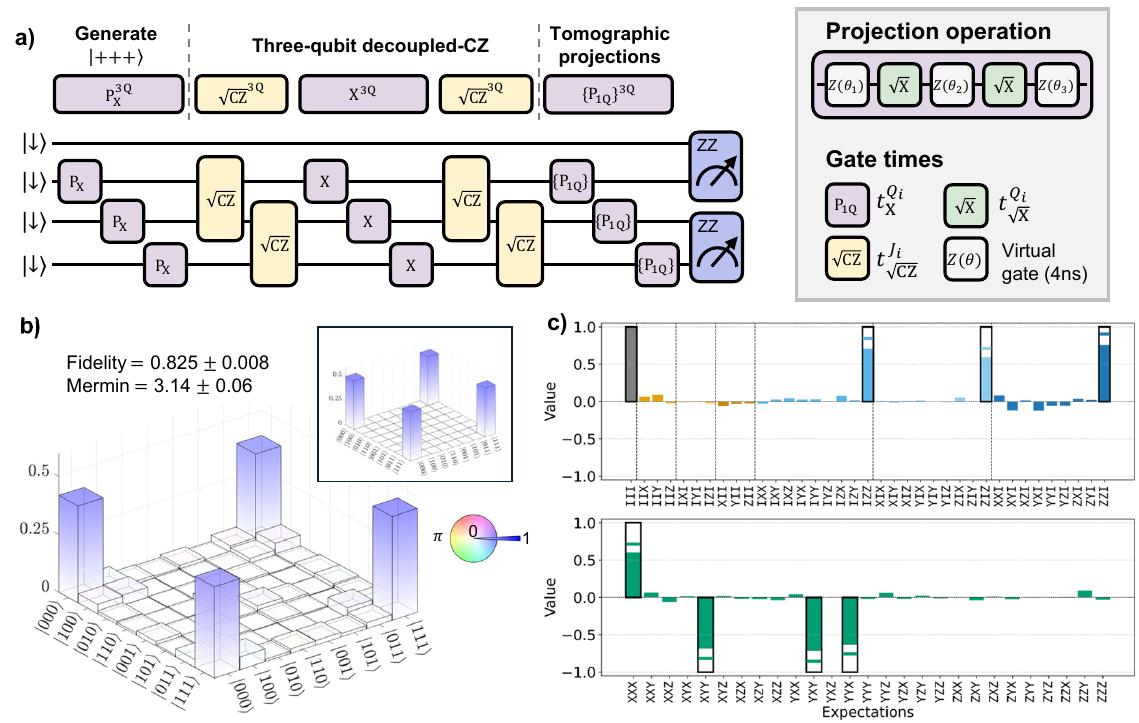}
    \caption{\textbf{Three-qubit entangled state measurements.}
    \textbf{a,} Circuit used to generate a three-qubit cluster state (equivalent to the GHZ state up to local single-qubit gates) and perform state tomography. $\{\textrm{P}_\textrm{1Q}\}$ are the set of single-qubit pre-measurement rotations required to perform state tomography. Two-qubit operations $\{\textrm{P}_\textrm{2Q}\}$ are also used on $\textrm{Q}_3$ and $\textrm{Q}_4$ to obtain $ZI$/$IZ$ measurements (see Extended Data Fig.~\ref{fig:extended_fig_6}b for circuit). 
    \textbf{b,} Reconstructed density matrix. A local basis transformation has been applied in post-processing to represent it as the well-recognised GHZ state density matrix. The SPAM error associated with dephasing from the tomographic operations ($\{\textrm{P}_\textrm{1Q}\}/\{\textrm{P}_\textrm{2Q}\}$) is compensated for in both the density matrix and quoted fidelity and Mermin value. The theoretical density matrix is shown in the insert.
    \textbf{c,} Measured Pauli operator expectation values with aforementioned local basis transformation applied. The filled bar represents the measured expectation value, the horizontal line is the estimated value with SPAM compensation, and the hollow black bars represent the theoretical values.
    For the density matrix and expectation values with no basis transformation or SPAM compensation see Extended Data Fig.~\ref{fig:extended_fig_7}a. 
    }
    \label{fig:main_fig_3}
\end{figure*}

\begin{figure*}[ht!]
    \includegraphics[width=\textwidth]{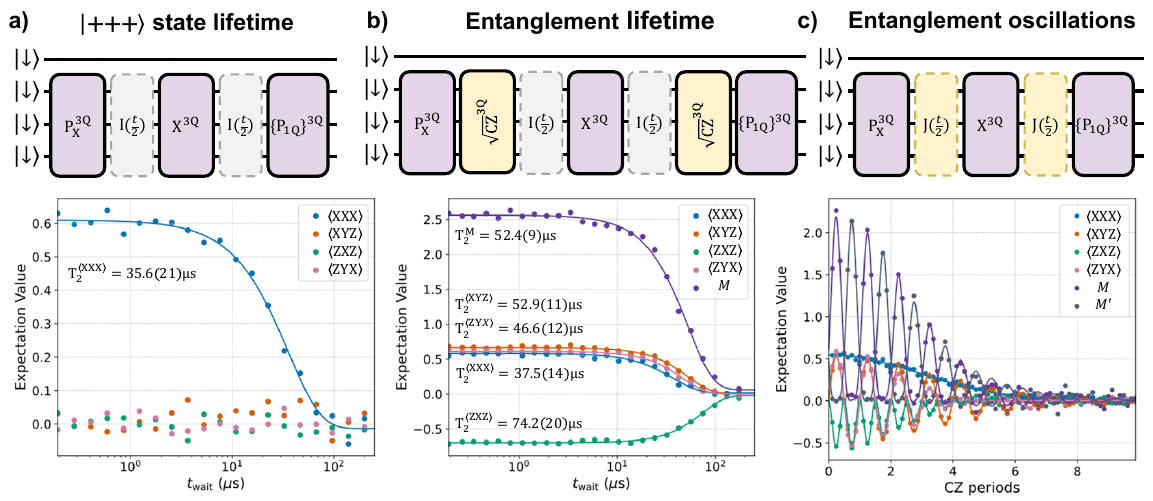}
    \caption{\textbf{Entanglement lifetime.}
    \textbf{a,} $T_2^\textrm{Hahn}$-style measurement on the three qubit state $\ket{\textrm{+++}}$ measured by removing the CZ operations from the circuit in Fig.~\ref{fig:main_fig_3}a and adding idle times. 
    \textbf{b,} Mermin value as function of idle time, representing the longevity of the entanglement between the three qubits. This is measured by inserting wait times on either side of the refocusing pulses.
    \textbf{c,} Measurement of the oscillations in Mermin value when the exchange time is extended. Here the exchange time for both J2 and J3 are increased proportionally with each other. 
    Error bounds on $T_2$ values represent 1$\sigma$ uncertainty.
    }
    \label{fig:main_fig_4}
\end{figure*}

\section*{Three qubit entanglement}

To characterise the overall performance of the processor, we investigate the degree of entanglement which can be generated between $\textrm{Q}_2$, $\textrm{Q}_3$ and $\textrm{Q}_4$. This includes benchmarking a cluster state and the GHZ state, and measuring entanglement lifetime. 

To begin, we use the circuit in Fig.~\ref{fig:main_fig_3}a to create a three-qubit cluster state $\ket{\psi}_\textrm{Cluster} = \frac{1}{\sqrt{2}}(\ket{i}\ket{0}\ket{i}-\ket{-i}\ket{1}\ket{-i})$. This is a maximally entangled state with uniform probability distribution over the computational basis. For the three-qubit case it is a GHZ-class state, equivalent to the GHZ state up to local single-qubit gates. Experimentally, the circuit used to generate this state can be designed in a laddered structure that allows each qubit to maintain symmetry with respect to dynamical decoupling pulses embedded in the dCZ gates. 

The circuit consists of single-qubit operations used to generate the state $\ket{\textrm{+++}}$, followed by a three-qubit dCZ operation which entangles the qubits, similar to that used in \cite{takeda_quantum_2021}. In order to reconstruct the state via tomography, pre-measurement rotations are applied prior to readout. These rotations are represented by the set $\{\textrm{P}_\textrm{1Q}\}$. We construct each gate in this set such that it consists of two $\sqrt{\textrm{X}}$ gates interleaved between three virtual phase gates (see Extended Data Fig.~\ref{fig:extended_fig_6}c and Methods for the full set of operations). Gates from this set are used for both the preparation of $\ket{\textrm{+++}}$ as well as the tomographic projections. Combining this gate implementation with the laddered structure of the circuit ensures each qubit has equal evolution time before and after refocusing. This enables Hahn-echo-like decoupling, which is important given that the total circuit time is comparable to $T_2^*$. 

An additional set of two-qubit pre-measurement rotations $\{\textrm{P}_\textrm{2Q}\}$ are also required for to obtain $ZI/IZ$ basis measurements from the $\textrm{Q}_3$-$\textrm{Q}_4$ pair. These gates combine a dCZ with operations from $\{\textrm{P}_\textrm{1Q}\}$ in order to maintain the same refocusing symmetry. See Extended Data Fig.~\ref{fig:extended_fig_6}b,d for details.

We perform 1000 single-shot measurements of each of the 360 combinations of tomographic pre-rotations and reconstruct the density matrix and the expectation values of the Pauli operators from the results (Fig.~\ref{fig:main_fig_3}b,c). We apply a local basis transformation in post-processing (see Methods) to convert the density matrix and Pauli operator expectations of the measured cluster state to that of the well-recognised GHZ state: $\frac{1}{\sqrt{2}}(\ket{000}+\ket{111})$.

Given the number of gates used in the $\textrm{P}_\textrm{1Q}$ and $\textrm{P}_\textrm{2Q}$ operations, a significant degree of dephasing is introduced by the measurement pre-rotations themselves. We treat this as a SPAM error and compensate for it using state tomography data taken on the initialised state $\ket{\downarrow\downarrow\downarrow}$ (see Methods for analysis). With compensation applied, we estimate a state fidelity of \SI{82.5(8)}{\percent}. We calculate the Mermin witness operator \cite{guhne_entanglement_2009} to be $M_\textrm{Cluster}=3.14(6)$. This violates the classical Mermin-witness bound, certifying multipartite entanglement \cite{mermin_extreme_1990}. The raw data and metrics with no transformation or SPAM compensation applied are shown in Extended Data Fig.~\ref{fig:extended_fig_6}a. It is noted that applying the single-qubit basis transformation in post-processing only reallocates the measured Pauli operator expectations to those associated with the GHZ state, and has no impact on the magnitude of the expectations, the degree of entanglement, or the measured state fidelity. 

We also measure the GHZ state directly by adding the required single-qubit gates into the circuit. The results are plotted in Extended Data Fig.~\ref{fig:extended_fig_7}b. We obtain a SPAM corrected state fidelity of \SI{74.6(9)}{\percent} and Mermin value of $M_\textrm{GHZ}=2.62(6)$. Without the correction these metrics are \SI{64.7(7)}{\percent} and $2.20(5)$ respectively. The measured fidelities for the cluster and GHZ states in this SiMOS device are comparable to initial three-qubit entanglement demonstrations in Si/SiGe quantum dots~\cite{takeda_quantum_2021, philips_universal_2022}, albeit lower than donor system demonstrations~\cite{madzik_precision_2022, thorvaldson_grovers_2025, edlbauer_11-qubit_2025} and recent work using shuttling~\cite{undseth_weight-four_2026}. 

The limiting factor in these circuits is the performance of the $\sqrt{\textrm{X}}$ gate and the associated spectator qubit dephasing as indicated by the $(\sqrt{X} \otimes I)$ GST fidelities of between \qtyrange{97}{98}{\percent}. Given that the GHZ circuit consists of 20  $\sqrt{X}$ gates (18 for the cluster state), the overall state fidelity degrades significantly. Additionally, the GHZ state circuit breaks the symmetry property due to the addition of the two extra gates, so will suffer from sub-optimal refocusing. 

Lastly, we characterise the entanglement lifetime in the three-qubit cluster state. As a reference experiment, we perform a Hahn-echo-like experiment on the state $\ket{\textrm{+++}}$ by removing the exchange pulses from our circuit, and inserting wait times before and after the refocusing pulses (see Fig.~\ref{fig:main_fig_4}a). By fitting the decay we measure $T_2^{\langle\textrm{XXX}\rangle} = $~\SI{35.6(21)}{\micro\second} due to the combined dephasing of all three qubits. 

Reinserting the $\sqrt{\textrm{CZ}}$ operations, we can calculate $M_\textrm{Cluster}$ as a function of wait time, as shown in Fig.~\ref{fig:main_fig_4}b. We first note that the decay rate of $\langle\textrm{XXX}\rangle$ compared to  Fig.~\ref{fig:main_fig_4}a has not been impacted by the addition of the $\sqrt{\textrm{CZ}}$ gates. Longer coherence times are measured on the other three Pauli operator expectations ($\langle\textrm{XYX}\rangle$, $\langle\textrm{ZXZ}\rangle$, $\langle\textrm{ZYX}\rangle$) because one or more or the constituent qubits are protected from dephasing by the eigenbasis. The coherence time of the overall entanglement, as measured by $M_\textrm{Cluster}$ is \SI{52.4(9)}{\micro\second}, corresponds closely to the average of the $T_2$ times of the four Pauli operator expectations. 

To test the impact of the entangling operations, the total exchange time between $\textrm{Q}_2$--$\textrm{Q}_3$ and $\textrm{Q}_3$--$\textrm{Q}_4$ is increased up to 10 full periods in Fig.~\ref{fig:main_fig_4}c. We calculate $M_\textrm{Cluster}$ as well as $M_\textrm{Cluster}'$, the Mermin witness operator for the cluster state $\ket{\psi}_\textrm{Cluster}'=\frac{1}{\sqrt{2}}(\ket{-i}\ket{0}\ket{-i}-\ket{i}\ket{1}\ket{i}$). $M_\textrm{Cluster}'$ is calculated with Equation~\ref{eqn:M_FES_alt} in Methods. $\ket{\psi}_\textrm{Cluster}'$ is generated when $\textrm{CZ}^\frac{3}{2}$ gates are applied rather than $\sqrt{\textrm{CZ}}$, as demonstrated in Extended Data Fig.~\ref{fig:extended_fig_8}. Hence in Fig.~\ref{fig:main_fig_4}c, periodic, out of phase maximisations in $M_\textrm{Cluster}$ and $M_\textrm{Cluster}'$ are observed with roughly 5 full oscillations before the state completely decoheres, corresponding 10 maximisations of entanglement. 

\section*{Outlook}

We demonstrate exchange coupling between two DQD unit cells of a SiMOS qubit array, enabling entanglement between three spin qubits. This is achieved through a narrow gate pitch and by operating the quantum dots at higher electron occupation, exploiting the larger spatial extent of p‑orbital wavefunctions. This is a meaningful step forward for the SiMOS platform, which has recently demonstrated the ability to fabricate larger arrays of unit cells~\cite{nickl_eight-qubit_2025, lim_2x2_2025}, but has so far been limited to operation within isolated unit cells.
 
Furthermore, we generalise previously developed methods for operating single unit cells to larger arrays of coupled DQDs, demonstrating that SPAM operations can be performed in parallel. We introduce a laddered gate sequence for generating entangled states which can be extended to arbitrarily many qubits while still maintaining optimally timed dynamical decoupling, independent of qubit‑to‑qubit variations in gate times. Using this circuit, we generate maximally entangled cluster and GHZ states, and show the coherence of this entanglement is extended up to a $T_2^\mathrm{Hahn}$ limit. By compensating for SPAM error we verify genuine tripartite entanglement.  
 
While the entangled state fidelities achieved in this academic device are modest, they are likely limited by charge noise given the \SI{50}{ppm} concentration of nuclear spin isotopes. Results from foundry‑fabricated devices report improved oxide uniformity and reduced charge noise~\cite{elsayed_low_2024, chittock-wood_radio-frequency_2025, loenders_understanding_2026}, along with improved single‑ and two‑qubit fidelities and $T_2$ times an order of magnitude longer than those observed here~\cite{zwerver_qubits_2022, steinacker_industry-compatible_2025, nickl_eight-qubit_2025}. Overall, these results provide a promising step toward scalable SiMOS spin‑qubit architectures, demonstrating both inter‑unit‑cell coupling, as well as control and operating techniques that scale favourably with qubit count.

\bibliographystyle{naturemag}
\bibliography{references}

\section*{Methods}

\subsection*{Measurement setup and cryogenics}
The device is wire-bonded to a custom designed and fabricated PCB, and housed in an enclosure mounted on the cold finger of the dilution refrigerator (a BlueFors LD400). All experiments were conducted at base temperature, and an American Magnetics AMI430 6-1-1 vector magnet was used to provide the DC magnetic field to the sample. A QDevil QDAC-I is used to provide DC voltage biasing. A Quantum Machines OPX+ was used to provide base-band voltage pulses to gate electrodes, operate the rf-SETs, and perform the real-time logic required for the initialisation state filter. Additionally, the OPX+ generates IQ modulation signals, which were input to a Keysight PSG8267D vector signal generator connected to the on-chip antenna, and used to perform single-qubit control. For further details on the measurement setup please refer to Methods and Extended Data of \cite{jones_mid-circuit_2026}. 

\subsection*{Gate set tomography}
To benchmark the gate performance and characterise the dominant error channels of the single- and two- qubit gate operations, a series of two qubit GST experiments were conducted on qubit pairs $\textrm{Q}_2$--$\textrm{Q}_3$, $\textrm{Q}_3$--$\textrm{Q}_4$ and $\textrm{Q}_2$--$\textrm{Q}_4$. In the case of the former two experiments, the gate sets characterised are: $\{\sqrt{\textrm{X}} \otimes \textrm{I}, \textrm{I} \otimes \sqrt{\textrm{X}}, \sqrt{\textrm{Z}} \otimes \textrm{I}, \textrm{I} \otimes \sqrt{\textrm{Z}}, \textrm{dCZ} \}$, where dCZ is a decoupled-CZ gate between the two qubits being characterised. For the  $\textrm{Q}_2$--$\textrm{Q}_4$ experiment, the gate set is limited to $\{\sqrt{\textrm{X}} \otimes \textrm{I}, \textrm{I} \otimes \sqrt{\textrm{X}}, \sqrt{\textrm{Z}} \otimes \textrm{I}, \textrm{I} \otimes \sqrt{\textrm{Z}} \}$, omitting the dCZ gate as we do not calibrate for direct exchange coupling between  $\textrm{Q}_2$ and $\textrm{Q}_4$. 

By performing this pair-wise GST on each combination of qubits, we are able to capture the idle dephasing and crosstalk errors on both spectator qubits during each single-qubit gate, without extending the subspace to 3 qubits, which would substantially increase the size of the GST experiment and required analysis time. 

\subsection*{Dual readout techniques}
Readout is performed in each qubit pair via parity mode PSB \cite{seedhouse_pauli_2021}. The readout sequence consists of an initial voltage ramp to a detuning point close to the charge transition region but still in the ($N_\text{P1}$,$N_\text{P2}$) = (3,5)/($N_\text{P3}$,$N_\text{P4}$) = (5,3) configuration. Here a reference signal is obtained by integrating the SET signal for \SI{100}{\micro\second}. The DQD is then detuned to the readout point in the PSB region with a \SI{20}{\nano\second} linear ramp, and a second integration of \SI{100}{\micro\second} occurs to obtain a read signal. The final readout signal which is the difference between the reference and read values. A threshold is applied to classify the measurement outcome as blockaded (even parity) or unblockaded (odd parity). 

In the case of the sequential readout technique, this readout sequence is executed while the neighbouring DQD is being held at the single-qubit control point in (3,5)/(5,3). Following readout, the DQD is ramped back from the PSB region to this single-qubit control point, and the readout of the second pair begins. In the case of the simultaneous readout sequence, both DQDs are pulsed to the reference and PSB detunings together, and the SET1 and SET2 signals are integrated in parallel. This reduces the total readout integration time required to measure both pairs from \SI{400}{\micro\second} to \SI{200}{\micro\second}.

\subsection*{Parallel readout calibration}
To determine a suitable detuning for both DQDs which enables parallel PSB readout, the experiment in Extended Data Fig.~\ref{fig:extended_fig_4}a is performed. Both qubit pairs are prepared in an equal superposition of odd and even parity states. The simultaneous readout sequence described above is then executed, however the detuning used in the read phase is stepped from one side of the PSB region to the other. After thresholding the readout we obtain the maps plotted in  Extended Data Fig.~\ref{fig:extended_fig_4}b.

The detunings used are selected such that both plots are in the PSB region, with care given not to bias the DQD in the cascaded readout regions outlined in Extended Data Fig.~\ref{fig:extended_fig_4}b. The suitable biasing region is easily seen by adding the two SET outputs together, as shown in Extended Data Fig.~\ref{fig:extended_fig_4}c. 

To verify that the selected biasing provides independent parity readout in both unit cells, the experiment in Extended Data Fig. \ref{fig:extended_fig_4}d is performed whereby Rabi oscillations are driven on all qubits followed by simultaneous readout of both pairs. The resulting parity maps are plotted in Extended Data Fig. \ref{fig:extended_fig_4}e. We see that in both unit cells the even (odd) parity states are map correctly blockaded (unblockaded) outcomes. Importantly, the Rabi oscillation frequencies observed in these maps correlate to the individually measured Rabi frequencies plotted in Fig.~\ref{fig:main_fig_1}c, verifying that the parity measurement in each unit cell is independent of the neighbouring unit cell. 

\subsection*{Heralded initialisation}
To enable high fidelity initialisation of the ground energy state across the array, we use a heralded initialisation protocol that involves real-time logic implemented by the FPGA. This procedure is as follows: 
\begin{enumerate}
    \item The array is held in the ($N_\text{P1}$,$N_\text{P2}$,$N_\text{P3}$,$N_\text{P4}$) = (4,4,4,4) charge configuration to ensure both DQDs relax to ground singlet state prior to initialisation. 
    \item The detuning in both pairs is ramped adiabatically, from the (4,4,4,4) charge configuration to (3,5,5,3). This ramp rate is tuned such that a high purity $\ket{\textrm{Q}_1, \textrm{Q}_2, \textrm{Q}_3, \textrm{Q}_4} = \ket{\downarrow\uparrow\uparrow\downarrow}$ state is populated. A fixed ramp rate is used for both DQDs, however the J1/J3 voltages applied when moving through the anti-crossing. This allows each DQD to be tuned independently to ensure $\ket{\textrm{Q}_1, \textrm{Q}_2} = \ket{\downarrow\uparrow}$ and $\ket{\textrm{Q}_3, \textrm{Q}_4} = \ket{\uparrow\downarrow}$ after ramped initialisation.  
    \item $X$ gates are applied to qubits $\textrm{Q}_2$ and $\textrm{Q}_3$ to convert the state to $\ket{\downarrow\downarrow\downarrow\downarrow}$
    \item Readout is performed in both DQD unit cells to measure Z-basis parity of both qubit pairs $M_\mathrm{ZZ}(\textrm{Q}_1, \textrm{Q}_2)$, $M_\mathrm{ZZ}(\textrm{Q}_3, \textrm{Q}_4)$. 
    \item The FPGA is used to execute a real-time logical condition. If even parity is measured in both unit cells ($M_\mathrm{ZZ}(\textrm{Q}_1, \textrm{Q}_2)=M_\mathrm{ZZ}(\textrm{Q}_3, \textrm{Q}_4)=0$), the initialisation is considered successful and the sequence progresses to quantum circuit execution. If either or both of the measurements register odd parity, $M_\mathrm{ZZ}=1$, the reinitialisation process is restarted from step 1, with both DQDs being ramped back to (4,4,4,4). 
\end{enumerate}
It is noted that this sequence does not distinguish the state preparation of $\ket{\downarrow\downarrow\downarrow\downarrow}$ from $\ket{\uparrow\uparrow\uparrow\uparrow}$. The likelihood of populating this state is very low (as verified with GST) owing to the purity of $\ket{\downarrow\uparrow\uparrow\downarrow}$ state after the ramped adiabatic initialisation. A secondary phase can be implemented however to distinguish these states, as performed in \cite{huang_high-fidelity_2024}.

While in this case the reinitialisation of both unit cells on unsuccessful initialisation attempts adds little overhead to the average state preparation time, it will become increasingly important to implement reinitialsation of individual unit cells as processors scale. 

\subsection*{Projection operations}
In order to perform state tomography, pre-measurement rotations must be applied to each qubit to obtain the orthogonal measurements required to reconstruct the state. We define these pre-rotations as "projection operations", $P_i^\textrm{Q}$, where $\textrm{Q}$ is the target qubit, and $i$ is the state projected to $+\textrm{Z}$ eigenstate by the operation. We construct these operations as hadamard style gates which effectively transform the basis. For example, $P_\textrm{X}^\textrm{Q}$ rotates $+\textrm{X}$ to $+\textrm{Z}$, and $-\textrm{X}$ to $-\textrm{Z}$, effectively converting a Z-basis measurement to an X-basis measurement. 

For $\textrm{Q}_2$, we use the set of projections $\{P_\textrm{1Q}^{\textrm{Q}_2}\}=\{ P_\textrm{X}^{\textrm{Q}_2}, P_\textrm{-X}^{\textrm{Q}_2}, P_\textrm{Y}^{\textrm{Q}_2}, P_\textrm{-Y}^{\textrm{Q}_2}, P_\textrm{Z}^{\textrm{Q}_2}, P_\textrm{-Z}^{\textrm{Q}_2} \}$. Given that $\textrm{Q}_1$ remains in $\ket{1}$, the parity PSB measurement between the two is effectively a state measurement of $\textrm{Q}_2$. 

For $\textrm{Q}_3$ and $\textrm{Q}_4$ we are required to extend our set of projection gates beyond $\{P_\textrm{1Q}\}$ to include the set of two-qubit operations $\{P_\textrm{2Q}\}$ listed in Extended Data Fig.~\ref{fig:extended_fig_6}d. These gates map the state of only one of the qubits into the parity of $\textrm{Q}_3$--$\textrm{Q}_4$, allowing for $ZI/IZ$ basis measurements rather than $ZZ$~\cite{leon_bell-state_2021, seedhouse_pauli_2021}. These are implemented using dCZ operations, and the same hadamard-style single-qubit pre-rotations are used instead of $\sqrt{\textrm{X}}$ and $\sqrt{\textrm{Y}}$, in order to maintain symmetric idle timing in the circuit. For each linearly independent pre-rotation listed in Extended Data Fig.~\ref{fig:extended_fig_6}d, we permute through both the positive and negative $\textrm{P}_\textrm{C}$ and $\textrm{P}_\textrm{D}$ projections, giving a total of 24 unique two-qubit projection operations. 

A total of 360 unique combinations of three qubit projection operations are performed ($6^3=216$ single-qubit rotations $\{P_\textrm{1Q}\}$, and $6\times24=144$ which incorporate the 2-qubit operations $\{P_\textrm{2Q}\}$ on $\textrm{Q}_3$--$\textrm{Q}_4$). 1000 repetitions of the target circuit are performed for each of these combinations, with the outcomes combined and averaged to obtain the expectation values of the 63 non-trivial Pauli operators. Linear inversion is used to reconstruct the density matrix. 

\subsection*{Density matrix transform}
The circuit in Fig.~\ref{fig:main_fig_3}a is designed to generate the cluster state $\ket{\psi}_\textrm{Cluster} = \frac{1}{\sqrt{2}}(\ket{i}\ket{0}\ket{i}-\ket{-i}\ket{1}\ket{-i})$. This has the density matrix: 
\begin{equation}
    \rho_\textrm{Cluster} = \frac{1}{8}
    \begin{bmatrix}
         1 & -i & -1 & -i & -i & -1 & -i & 1 \\
         i & 1 & -i & 1 & 1 & -i & 1 & i \\
         -1 & i & 1 & i & i & 1 & i & -1 \\
         i & 1 & -i & 1 & 1 & -i & 1 & i \\
         i & 1 & -i & 1 & 1 & -i & 1 & i \\
         -1 & i & 1 & i & i & 1 & i & -1 \\
         i & 1 & -i & 1 & 1 & -i & 1 & i \\
         1 & -i & -1 & -i & -i & -1 & -i & 1 \\
    \end{bmatrix}
    \label{rho_FES}
\end{equation}
This is represented by the matrix plotted in the insert of Extended Data Fig.~\ref{fig:extended_fig_7}a. After reconstructing the measured density matrix, $\hat{\rho}_\textrm{Cluster}$ (plotted in Fig.~\ref{fig:extended_fig_7}a), we transform the matrix to that plotted in Fig.~\ref{fig:main_fig_3}b, by applying the operator: 
\begin{equation}
    U\hat{\rho}_\textrm{Cluster} = \bigl( \sqrt{X} \otimes I \otimes \sqrt{X} \bigl) \hat{\rho}_\textrm{Cluster}
\end{equation}
where $I$ is the 2x2 identity. With the transform applied, the target matrix corresponds to GHZ state, as plotted in the insert in Fig.~\ref{fig:main_fig_3}b: 
\begin{equation}
    \rho_{\textrm{GHZ}} = \frac{1}{2}
    \begin{bmatrix}
        1 & 0 & 0 & 0 & 0 & 0 & 0 & 1 \\
        0 & 0 & 0 & 0 & 0 & 0 & 0 & 0 \\
        0 & 0 & 0 & 0 & 0 & 0 & 0 & 0 \\
        0 & 0 & 0 & 0 & 0 & 0 & 0 & 0 \\
        0 & 0 & 0 & 0 & 0 & 0 & 0 & 0 \\
        0 & 0 & 0 & 0 & 0 & 0 & 0 & 0 \\
        0 & 0 & 0 & 0 & 0 & 0 & 0 & 0 \\
        1 & 0 & 0 & 0 & 0 & 0 & 0 & 1 \\
    \end{bmatrix}
    \label{rho_GHZ}
\end{equation}

This also remaps the Pauli operator expectation values. For those relevant to the Mermin value, the following mapping occurs: 
\begin{equation}
    \begin{split}
        \langle\textrm{XXX}\rangle &\rightarrow \langle\textrm{XXX}\rangle \\
        \langle\textrm{XYZ}\rangle &\rightarrow -\langle\textrm{XYY}\rangle \\
        \langle\textrm{ZXZ}\rangle &\rightarrow \langle\textrm{YXY}\rangle \\
        \langle\textrm{ZYX}\rangle &\rightarrow -\langle\textrm{YYX}\rangle \\ 
    \end{split}
    \label{eqn:M_expectations_transformation}
\end{equation}

\subsection*{Mermin value calculations}
We use the Mermin witness function \cite{guhne_entanglement_2009} to probe the degree of entanglement in our generated state. For the GHZ state $\frac{1}{\sqrt{2}}(\ket{000} + \ket{111})$ the Mermin value is defined as:
\begin{equation}
    M_\textrm{GHZ} = \langle \textrm{XXX}\rangle - \langle \textrm{XYY}\rangle - \langle \textrm{YXY}\rangle -\langle \textrm{YYX}\rangle.
    \label{eqn:M_GHZ}
\end{equation}

If we consider the cluster state plotted in Extended Data Fig.~\ref{fig:extended_fig_7}a, the Pauli operator expectations are rotated with respect to those of used in equation~\ref{eqn:M_GHZ} above, under the mapping defined by equation~\ref{eqn:M_expectations_transformation}. Hence, for the cluster state, the equivalent Mermin value is calculated as: 
\begin{equation}
    M_\textrm{Cluster} = \langle \textrm{XXX}\rangle + \langle \textrm{XYZ}\rangle - \langle \textrm{ZXZ}\rangle + \langle \textrm{ZYX}\rangle.
    \label{eqn:M_FES}
\end{equation}

As explained in the main text, if we extend the time that the exchange interaction is on such that a $\textrm{CZ}^{\frac{3}{2}}$ gate is performed rather than $\textrm{CZ}^{\frac{1}{2}}$, an alternative cluster state $\ket{\psi}_\textrm{Cluster}'=\frac{1}{\sqrt{2}}(\ket{-i}\ket{0}\ket{-i}-\ket{i}\ket{1}\ket{i})$ is created. This has density matrix: 
\begin{equation}
    \rho'_\textrm{Cluster}= \frac{1}{8}
    \begin{bmatrix}
         1 & i & -1 & i & i & -1 & i & 1 \\
         -i & 1 & i & 1 & 1 & i & 1 & -i \\
         -1 & -i & 1 & -i & -i & 1 & -i & -1 \\
         -i & 1 & i & 1 & 1 & i & 1 & -i \\
         -i & 1 & i & 1 & 1 & i & 1 & -i \\
         -1 & -i & 1 & -i & -i & 1 & -i & -1 \\
         -i & 1 & i & 1 & 1 & i & 1 & -i \\
         1 & i & -1 & i & i & -1 & i & 1 \\
    \end{bmatrix}
    \label{rho_FES_alt}
\end{equation}

Like the initial cluster state, this is a maximally entangled state with uniform probability distribution across the computational basis. However, compared to $\ket{\psi}_\textrm{Cluster}$, the entanglement in $\ket{\psi}_\textrm{Cluster}'$ is encoded in alternate phase relations between the constituent qubits. The relavent Pauli operators for the Mermin values remain the same, however due to the phase shift the sign of the $\langle\textrm{XYZ}\rangle$ and $\langle\textrm{ZYX}\rangle$ are flipped: 
\begin{equation}
    M_\textrm{Cluster}' = \langle \textrm{XXX}\rangle - \langle \textrm{XYZ}\rangle - \langle \textrm{ZXZ}\rangle - \langle \textrm{ZYX}\rangle.
    \label{eqn:M_FES_alt}
\end{equation}

The relationship between the exchange times, the expectations values of the four Pauli operators of interest, and the calculated $M_\textrm{Cluster}$ and $M_\textrm{Cluster}'$ can be seen in Extended Data Fig.~\ref{fig:extended_fig_8}.

\subsection*{SPAM error correction}
To obtain a more accurate estimate of the fidelity of the measured entangled states, we compensate for dephasing introduced by the tomographic projection operators $\textrm{P}_{\textrm{1Q}}$ and $\textrm{P}_{\textrm{2Q}}$. To isolate the effect of these operations, we remove the $\textrm{P}_\textrm{X}$ rotations and dCZ gates gates from the circuit in Fig.~\ref{fig:main_fig_3}a, and perform full state tomography on the initial state $\ket{\downarrow\downarrow\downarrow}$. The results, shown in Extended Data Fig.~\ref{fig:extended_fig_7}c, indicate a reduction in the measured state fidelity compared to the GST-derived initialisation and readout fidelities, demonstrating that these operations introduce decoherence. We treat this decoherence as a contribution to the SPAM error.

We model the SPAM error as a uniform depolarising channel acting on the three-qubit Hilbert space. Under this assumption, the effect can be described as a global contraction of the Bloch vector. The contraction factor is estimated from the measured purity of the reconstructed initial state,
\begin{equation}
    \lambda = \sqrt{\frac{\textrm{Tr}(\hat{\rho_0}^2) - 1/d}{1-1/d}}\,,
\end{equation}
where $\hat{\rho}_0$ is the density matrix obtained from tomography of the initial state, and $d = 8$ is the Hilbert space dimension.

The measured Pauli operator expectation values of the entangled states are then rescaled by $1/\lambda$, excluding the identity component $\langle III \rangle$. A SPAM-corrected density matrix is reconstructed from these rescaled expectation values and used to extract corrected estimates of the state fidelity and Mermin witness. 

While this approach does not constitute a complete characterization of SPAM errors, it provides a physically motivated approximation of the dominant dephasing errors imparted by the tomographic projection operations, enabling a more faithful estimate of the generated entangled states.

\subsection*{Data acquisition and processing}

Data acquisition is done via the OPX+ input channel. Measurement shots are processed during the measurement through the FPGA on the OPX+, including sensor current thresholding and real-time logic. The measurement result is then uploaded to the server. The data are processed in Python.
 
\raggedbottom
\section*{Data availability}
All data generated or analyzed during this study are available from the corresponding author upon reasonable request.

\section*{Code availability}
Analysis and control code are available upon reasonable request.

\section*{Author contributions}
\noindent
C.J. conducted the experiments with assistance from J.Y.H, and under the supervision of C.H.Y.
Additional input was given by M.K.F., S.S., P.S., T.T., W.H.L., A.S.D., A.S., and A.L.\,.
S.S. conducted the initial device screening and characterisation. 
W. H. L., and F. E. H. designed and fabricated the devices.
N.V.A., H.-J.P., and M.L.W.T. provided the purified silicon substrate.
C.J., M.K.F., G.A.P. and C.H.Y. performed analysis of experimental data.
C.J. wrote the manuscript, with contributions from all authors.

\section*{Competing interests}
A.S.D. is CEO and a director of Diraq Pty Ltd. S.S., M.K.F., G.A.P., T.T., P.S., F.E.H., W.H.L., A.S.D., A.S., A.L., and C.H.Y. declare equity interest in Diraq. Other authors declare no competing interest.
\\
\\
\begin{acknowledgments}
We acknowledge technical support from A. Dickie. We acknowledge support from the Australian Re search Council (Grants No. FL190100167 and No. CE170100012), the U.S. Army Research Office (Grants No. W911NF-23-10092), the U.S. Air Force Office of Scientific Research (AFOSR) (Grants No. FA2386-22 1-4070), and the NSW Node of the Australian National Fabrication Facility. The views and conclusions contained in this document are those of the authors and should not be interpreted as representing the official policies, either expressed or implied, of the Army Research Office or the US Government. This research includes computations using the computational cluster Katana supported by Research Technology Services at UNSW Sydney. C.J. acknowledges support from Sydney Quantum Academy. This research was undertaken with the assistance of resources from the National Computational Infrastructure (NCI Australia), an NCRIS enabled capability supported by the Australian Government.

\end{acknowledgments}

\setcounter{figure}{0}
\setcounter{table}{0}
\captionsetup[figure]{name={\bf{Extended Data Fig.}},labelsep=line,justification=centerlast,font=small,singlelinecheck=false}
\captionsetup[table]{name={\bf{Extended Data Table}},labelsep=line,justification=centerlast,font=small,position=above}

\section*{Extended data}

\begin{figure*}[p]
    \includegraphics[width=\textwidth]{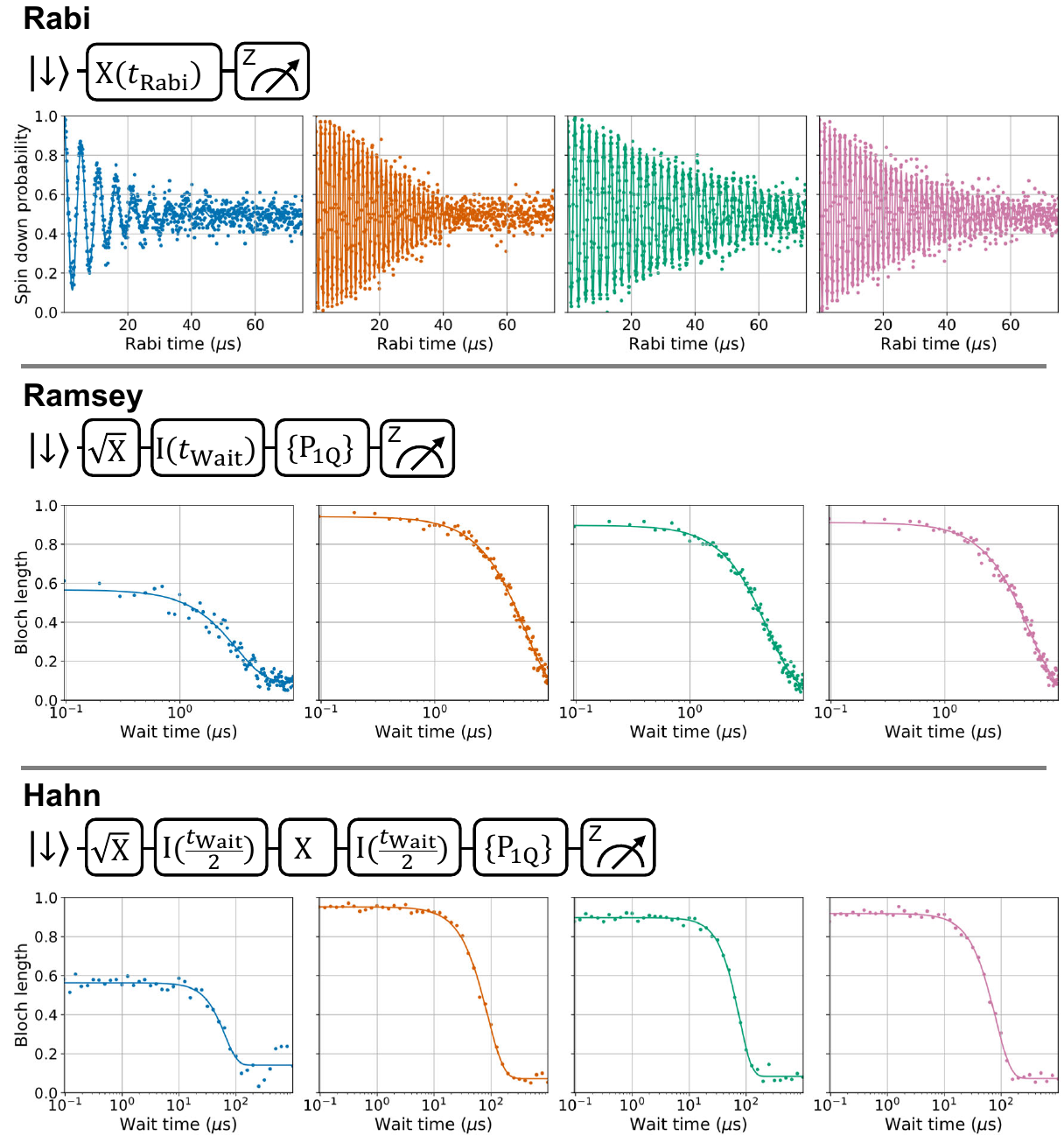}
    \caption{\textbf{Coherence time measurements.} Fitted coherence time data for Rabi, Ramsey and Hahn sequences used to obtain $T_\textrm{2}$ estimates in Extended Data Table \ref{tab:coherence_times}. In the case of the Ramsey and Hahn experiments, the measurement is performed for all six single-qubit projections $\{\textrm{P}_\textrm{1Q}\}$. From these projections we can calculate the Bloch length $\sqrt{\langle\sigma_\textrm{x}\rangle^2 +\langle\sigma_\textrm{y}\rangle^2 + \langle\sigma_\textrm{z}\rangle^2 }$, which is used to fit the decay. 
    }
    \label{fig:extended_fig_2}
\end{figure*}

\begin{center}
\captionof{table}{\textbf{Single-qubit metrics.} Extracted Rabi frequencies and $T_2$ coherence times for qubits $\textrm{Q}_1$–$\textrm{Q}_4$, obtained from the fitted data in extended data figure \ref{fig:extended_fig_2}. Uncertainties are given in parentheses correspond to $2\sigma$ error.}
\label{tab:coherence_times}
\begin{tabular}{|c|c|c|c|c|}
    \hline
    \textbf{Measurement} & \textbf{$\textrm{Q}_1$} & \textbf{$\textrm{Q}_2$} & \textbf{$\textrm{Q}_3$} & \textbf{$\textrm{Q}_4$} \\
    \hline
    $f_\textrm{Rabi}$ (kHz) 
        & 183.2(8) & 680.9(4) & 442.5(2) & 624.1(3) \\        
    $\mathrm{T_2^{Rabi}}$ ($\mu$s) 
        & 13.6(18) & 29.2(10) & 45.9(20) & 32.4(19) \\
    $\mathrm{T_2^*}$ ($\mu$s) 
        & 3.1(2) & 6.2(5) & 4.8(2) & 5.5(3) \\
    $\mathrm{T_2^{Hahn}}$ ($\mu$s) 
        & 64.0(104) & 87.2(34) & 76.3(29) & 79.4(34) \\
    \hline
\end{tabular}
\end{center}

\begin{center}
\captionof{table}{\textbf{Single-qubit gate fidelity.} All fidelity estimates obtained using GST, and correspond to $\sqrt{X}$ on the target qubit. Equivalent single-qubit on-target infidelities calculated, as well as the two-qubit fidelities $F_{\sqrt{X} \otimes I}$. These $F_{\sqrt{X} \otimes I}$ fidelities include the decoherence errors on the idle qubit while the target $\sqrt{X}$ is performed, which are not reflected in the on-target fidelities. Uncertainties are given in parentheses correspond to $2\sigma$ error.}
\label{tab:GST_1Q_fid}
\begin{tabular}{|c | c | c c c|}
    \hline
    \textbf{Target} & \textbf{On-target} & \multicolumn{3}{c|}{\textbf{Idle Qubit}} \\
    \cline{3-5}
    \textbf{qubit} & \textbf{fidelity}  & \textbf{$\textrm{Q}_2$} & \textbf{$\textrm{Q}_3$} & \textbf{$\textrm{Q}_4$} \\
    \hline
    $\textrm{Q}_2$ & \SI{99.78(37)}{\percent} & \cellcolor{lightgray} & \SI{97.36(16)}{\percent} & \SI{97.56(15)}{\percent} \\
    $\textrm{Q}_3$ & \SI{99.24(34)}{\percent}  & \SI{98.13(16)}{\percent}  & \cellcolor{lightgray} & \SI{97.56(20)}{\percent} \\
    $\textrm{Q}_4$ & \SI{99.14(32)}{\percent}  & \SI{97.99(12)}{\percent}  & \SI{97.35(21)}{\percent}  & \cellcolor{lightgray} \\
    \hline
\end{tabular}
\end{center}

\begin{center}
\captionof{table}{\textbf{Decoupled-CZ gate fidelity.} Two-qubit decoupled-CZ fidelities obtained using GST. Uncertainties are given in parentheses correspond to $2\sigma$ error.}
\label{tab:GST_2Q_fid}
\begin{tabular}{| c | c |}
    \hline
    \textbf{Qubit pair} & \textbf{Fidelity} \\
    \hline
    $\textrm{Q}_2$-$\textrm{Q}_3$ & \SI{95.82(19)}{\percent} \\
    $\textrm{Q}_3$-$\textrm{Q}_4$ & \SI{95.53(24)}{\percent} \\
    \hline
\end{tabular}
\end{center}

\begin{figure*}[p]
    \includegraphics[width=\textwidth]{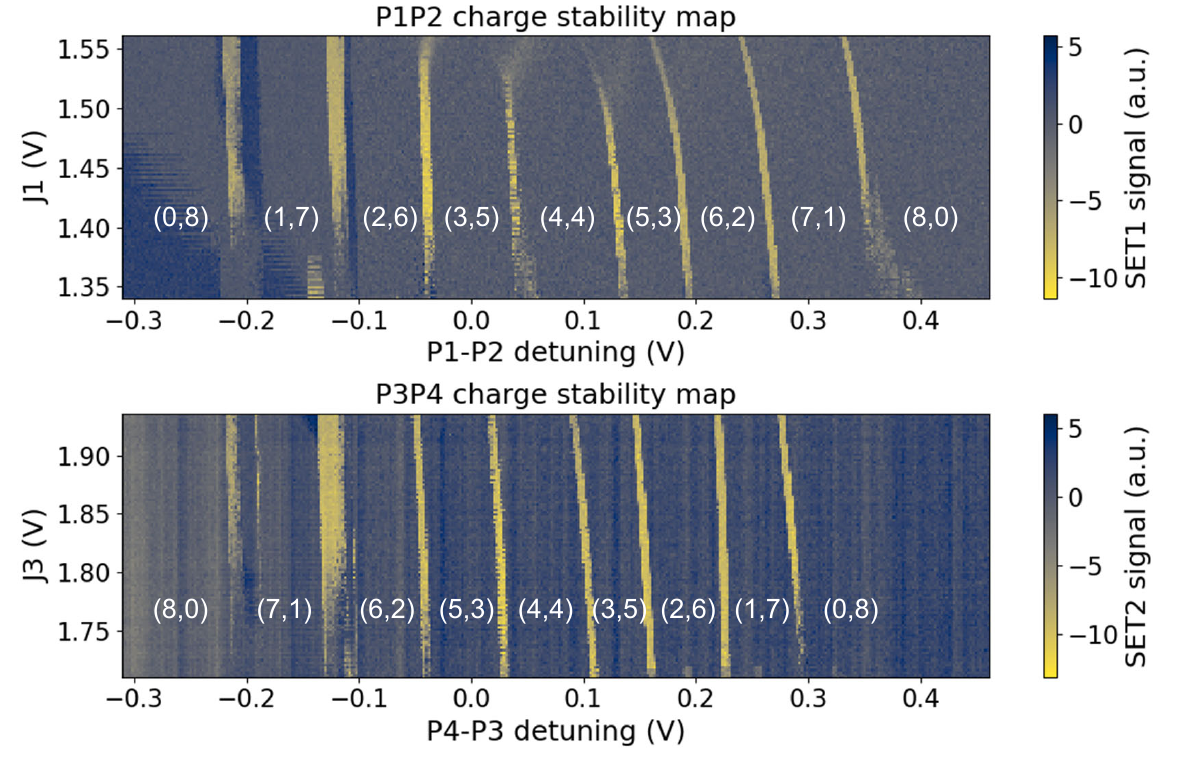}
    \caption{\textbf{Charge stability maps.} Charge stability diagram of DQDs P1-P2 (top) and P3-P4 (bottom) measured in isolated mode via SET1 and SET2 respectively. A higher amplitude voltage excitation signal is used to measure the leftmost two transitions, resulting in appearance of wider charge signals. labels correspond to the ($N_\textrm{P1},N_\textrm{P2}$) and ($N_\textrm{P3},N_\textrm{P4}$) charge regions in the top and bottom plots respectively. The nominal qubit regime is  ($N_\textrm{P1},N_\textrm{P2},N_\textrm{P3},N_\textrm{P4}$) = ($3,5,5,3$), with readout occuring towards the (4,4) transition in both DQDs.
    }
    \label{fig:extended_fig_1}
\end{figure*}

\begin{figure*}[p]
    \includegraphics[width=0.6\textwidth]{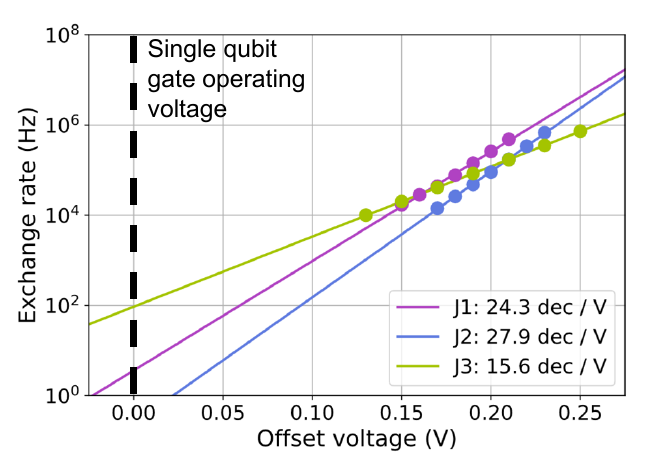}
    \caption{\textbf{Exchange rate controllability.} Exchange rate between $\textrm{Q}_1$-$\textrm{Q}_2$, $\textrm{Q}_2$-$\textrm{Q}_3$  and $\textrm{Q}_3$-$\textrm{Q}_4$ as a function of the voltage offsets $\Delta V_\textrm{J}$ on exchange gates J1, J2 and J3 respectively. Exchange rates are determined by fitting exchange oscillation measured in dCZ experiments performed at each offset voltage. Controllability is fitted to $J=a\exp({b\Delta V_\textrm{J}})+c$. Offset voltage of 0V represents the J-gate voltages at which single-qubit gates are performed.   
    }
    \label{fig:extended_fig_3}
\end{figure*}

\begin{figure*}[p]
    \includegraphics[width=\textwidth]{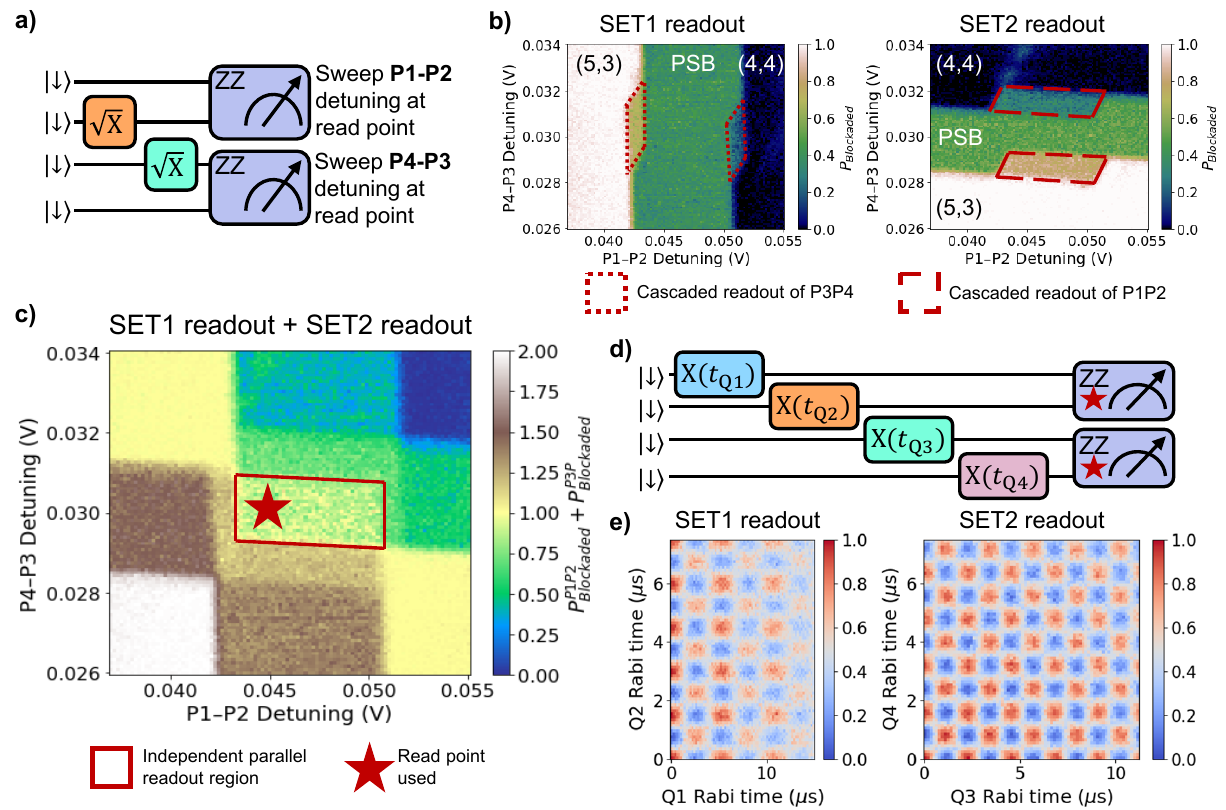}
    \caption{\textbf{Parallel readout calibration and verification.} 
    \textbf{a),} Measurement sequence used to determine biasing for parallel readout. See Methods for further details. 
    \textbf{b),} PSB regions in the P1P2 (left) and P3P4 (right) DQDs, as a function of the P1-P2 detuning ($\Delta V_\textrm{P1} = - \Delta V_\textrm{P2}$) on the x-axis and P3-P4 detuning ($\Delta V_\textrm{P4} = - \Delta V_\textrm{P3}$) on the y-axis. The dotted regions correspond to cascaded readout of the neighbour DQD. Parallel PSB occurs when both DQDs are biased to the PSB (green) regions. The gradients in the vertical (P1-P2) and horizontal (P3-P4) transitions are due to gate cross-capacitance. We observe discrete jumps in the PSB regions at $\Delta V_\textrm{P4} = - \Delta V_\textrm{P3} = 0.0285$ and $0.032$ in the P1-P2 readout, and $\Delta V_\textrm{P1} = - \Delta V_\textrm{P2} = 0.043$ and $0.052$ in the P3-P4 readout. These jumps arise from a change in charge confirguation in one of the DQDs, which shifts the potential of the neighbouring DQD. 
    \textbf{c),} Sum of the SET1 and SET2 thresholded signals from data in \textbf{b}. The outlined region in centre with value of 1 indicates the parallel parity readout region. The red star indicated the read point biasing used. 
    \textbf{d),} Sequence used to verify the independent parallel parity readout is achieved at biasing in \textbf{c}. Rabis are driven sequentially in each qubit, ensuring all parity pairs $\{ \ket{\downarrow\downarrow},\ket{\uparrow\uparrow}, \ket{\uparrow\downarrow}, \ket{\downarrow\uparrow} \}$ are measured. 
    \textbf{e),} Resulting parity map from the experiment in \textbf{d}. We observe the expected parity outcomes for each input state, and the oscillation frequencies match the Rabi frequencies of each individual qubit measured in Fig.~\ref{fig:main_fig_3}c. 
    }
    \label{fig:extended_fig_4}
\end{figure*}

\begin{figure*}[p]
    \includegraphics[width=\textwidth]{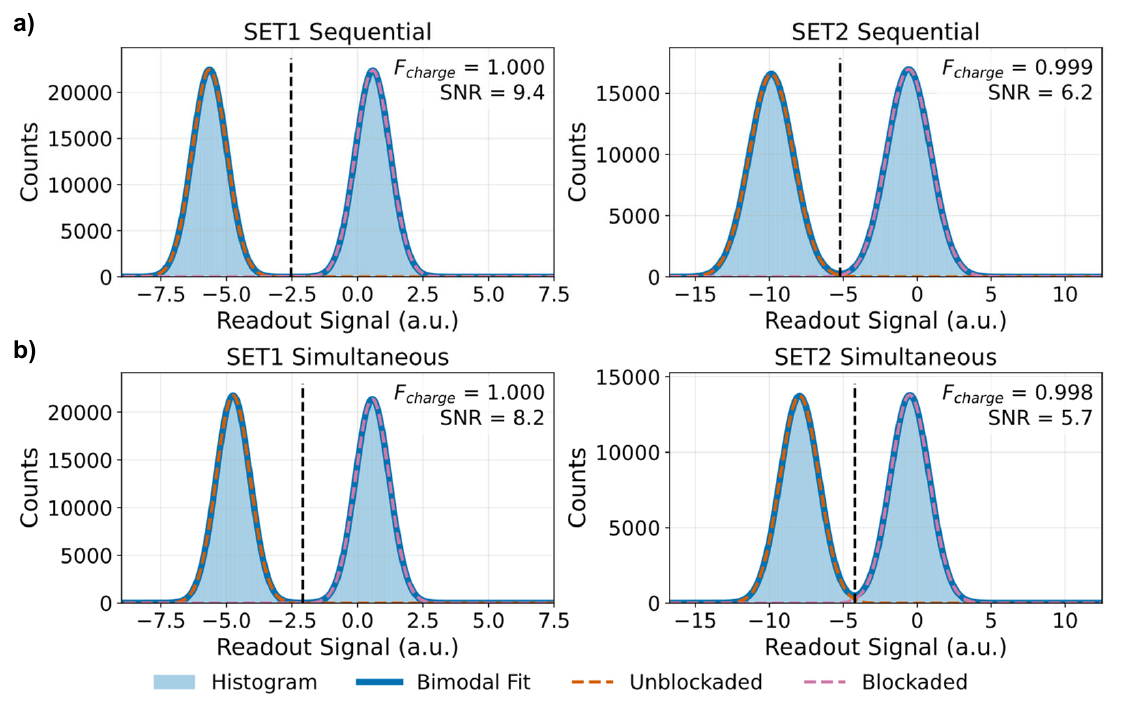}
    \caption{\textbf{Readout histogram analysis.} \textbf{a),} 1D histogram plots of sequential readout data in Fig~\ref{fig:main_fig_2}b. Bimodal Gaussian fitting is performed to extract the SNR and charge fidelity. 
    \textbf{b),} 1D histogram plots of simultaneous readout data in Fig~\ref{fig:main_fig_2}d.
    }
    \label{fig:extended_fig_5}
\end{figure*}

\begin{figure*}[p]
    \includegraphics[width=\textwidth]{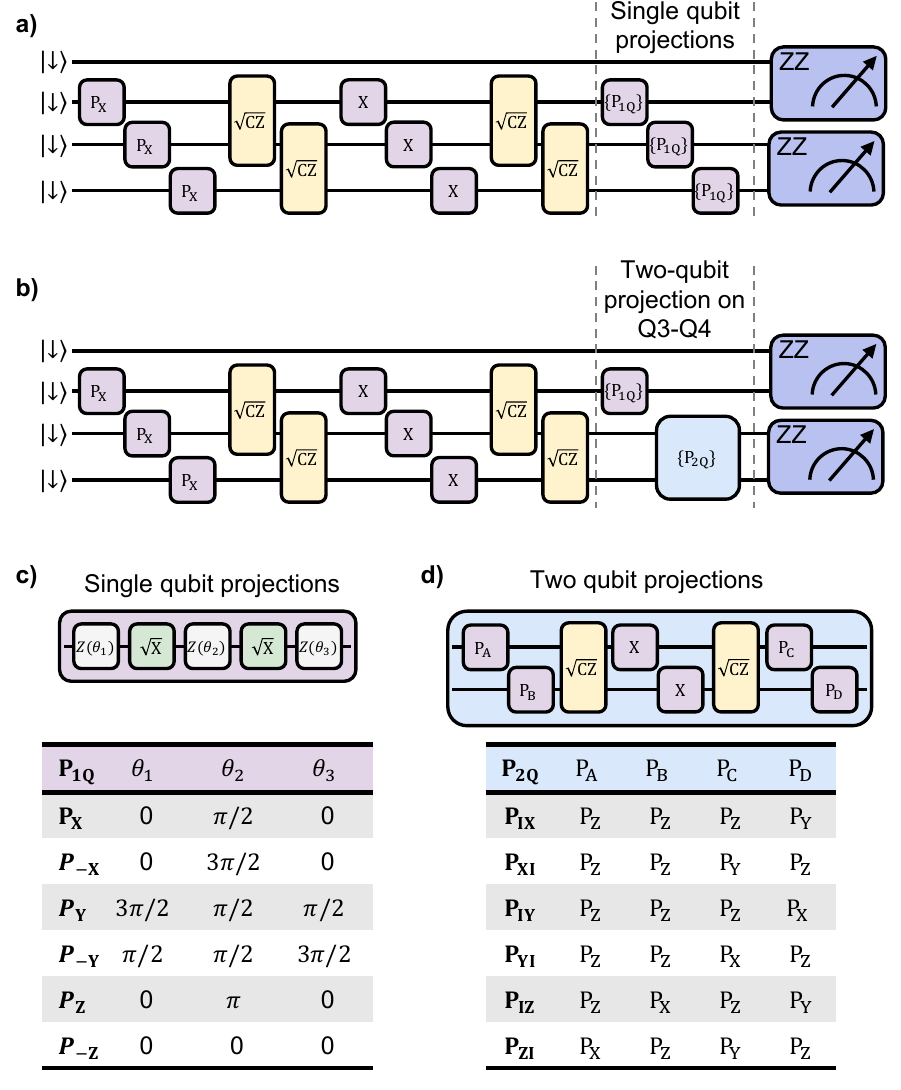}
    \caption{\textbf{Single- and two- qubit projection operations.} \textbf{a),} Circuit used to perform single-qubit projections $\{\textrm{P}_\textrm{1Q}\}$ on 3-qubit entangled state. 
    \textbf{b),} Circuit used to perform single-qubit projection $\{\textrm{P}_\textrm{1Q}\}$ on $\textrm{Q}_2$, and two-qubit projection $\{\textrm{P}_\textrm{2Q}\}$ on the $\textrm{Q}_3$-$\textrm{Q}_4$ pair..
    \textbf{c),} Implementation details of single-qubit projections $\{\textrm{P}_\textrm{1Q}\}$. 
    \textbf{d),} Implementation details of two-qubit projections $\{\textrm{P}_\textrm{2Q}\}$. 
    }
    \label{fig:extended_fig_6}
\end{figure*}

\begin{figure*}[p]
    \includegraphics[width=0.90\textwidth]{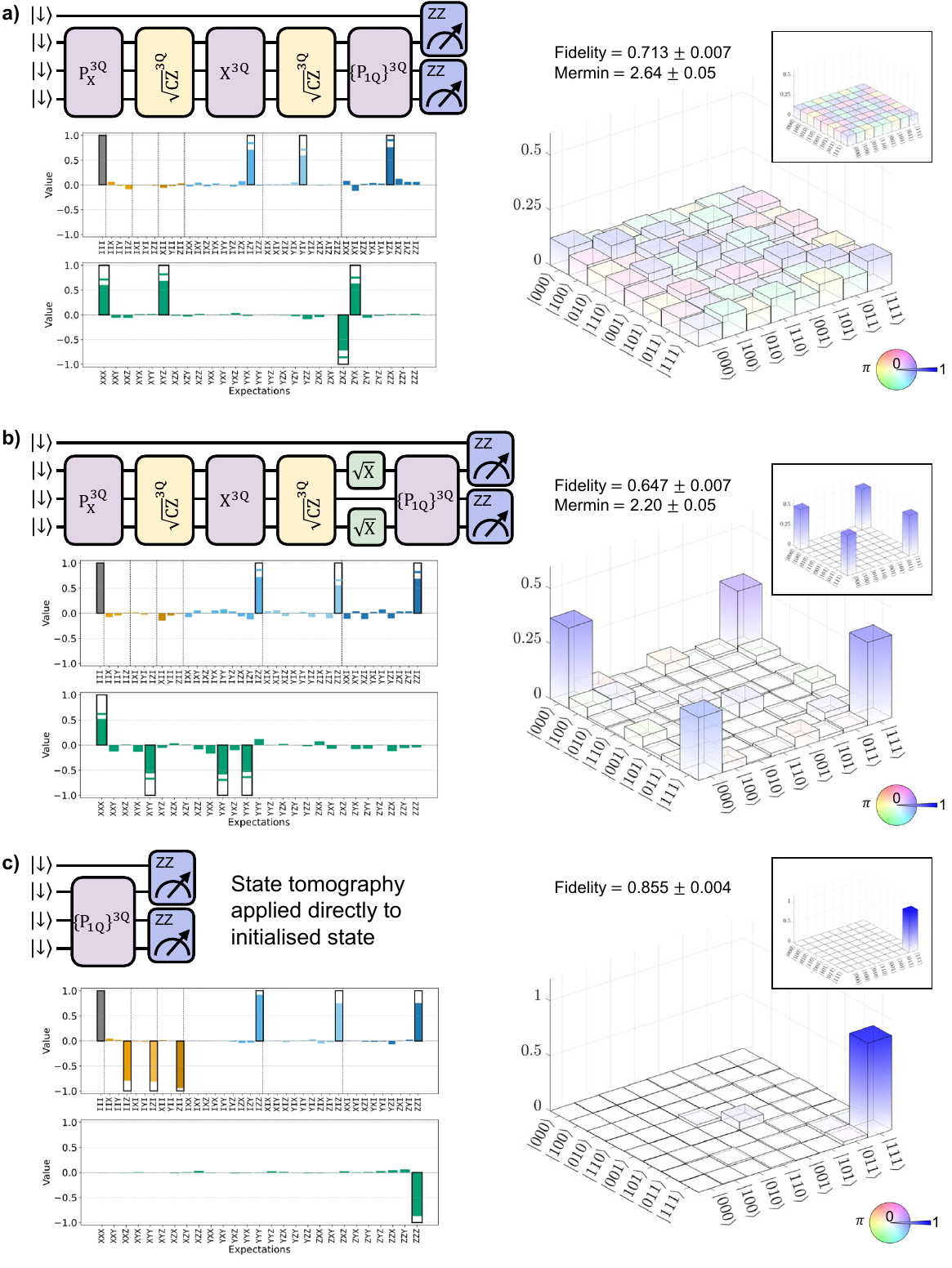}
    \caption{\textbf{Tomography of three-qubit states.}
    Experimental circuits, measured Pauli operator expectations, and reconstructed density matrices for
    \textbf{a),} Cluster state: $\ket{\psi}_\textrm{Cluster} = \frac{1}{\sqrt{2}}(\ket{i}\ket{0}\ket{i}-\ket{-i}\ket{1}\ket{-i})$;
    \textbf{b),} GHZ state: $\ket{\psi}_\textrm{GHZ} = \frac{1}{\sqrt{2}}(\ket{000}+\ket{111})$;
    \textbf{c),} Initialised state: $\ket{\psi}_0 = \ket{111}$.
    No local basis transformation, or compensation for SPAM has been applied to data in this figure, or the quoted fidelities and Mermin values.
    }
    \label{fig:extended_fig_7}
\end{figure*}

\begin{figure*}[p]
    \includegraphics[width=0.8\textwidth]{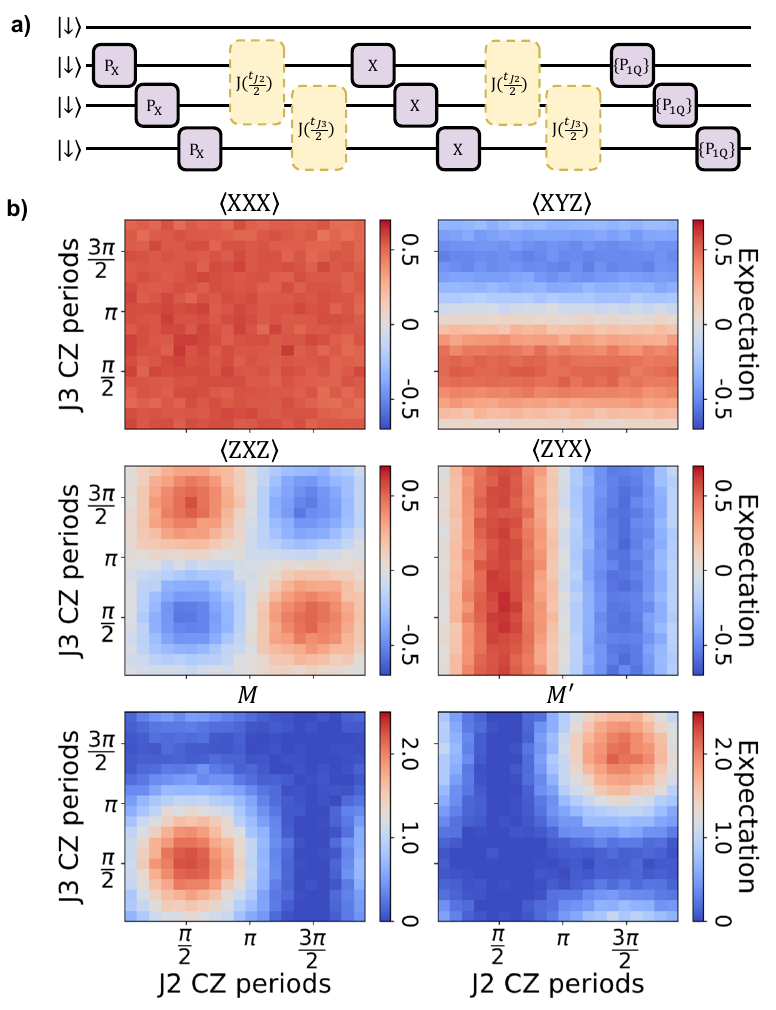}
    \caption{\textbf{Entangled vs CZ-time.} 
    \textbf{a),} Circuit used to measure oscillations in the Pauli operator expectation values relevant to the calculation of $M_\textrm{Cluster}$ and $M_\textrm{Cluster}'$ The wait times used in each exchange pulse, $t_\textrm{J2}$ and $t_\textrm{J3}$ are varied independently up to one oscillation period.  
    \textbf{b),} Plots of the measured expectation values of $\langle\textrm{XXX}\rangle$, $\langle\textrm{XYZ}\rangle$, $\langle\textrm{ZXZ}\rangle$ and $\langle\textrm{YZX}\rangle$, as well as the calculated Mermin  and alternate Mermin values, as a function of $t_\textrm{J2}$ and $t_\textrm{J3}$. Entanglement is maximised at 1/4 and 3/4 periods of the CZ oscillation. 
    }
    \label{fig:extended_fig_8}
\end{figure*}

\end{document}